\documentclass[journal=aamick,layout=draft]{achemso}
\setcitestyle{square}
\usepackage[version=4]{mhchem}
\usepackage[hyperindex]{hyperref}
\hypersetup{colorlinks,citecolor=blue,linkcolor=blue,urlcolor=blue}
\usepackage{units}
\usepackage{amsmath}
\usepackage{booktabs}
\usepackage{etoolbox}
\usepackage{xspace}
\SectionNumbersOff

\newcommand{\diffd}{\,\text{d}}
\newcommand{\voc}{V_\text{OC}}
\newcommand{\jsc}{J_\text{SC}}

\newcommand{\dopant}{\ce{Mo(tfd{-}CO2Me)3}\xspace}
\newcommand{\crosslinker}{benzene-1,3,5-triyl tris(4-azido-2,3,5,6-tetra\-fluoro\-benzo\-ate)\xspace}

\makeatletter
\patchcmd{\acs@contact@details}{E}{*\,E}{}{}
\makeatother

\author{Staffan Dahlström}
\email{stdahlst@abo.fi}
\affiliation{Physics, Faculty of Science and Engineering, \AA{}bo Akademi University, Porthansgatan 3, 20500 Turku, Finland}

\author{Sebastian Wilken}
\affiliation{Physics, Faculty of Science and Engineering, \AA{}bo Akademi University, Porthansgatan 3, 20500 Turku, Finland}

\author{Yadong Zhang}
\affiliation{School of Chemistry \& Biochemistry, Georgia Institute of Technology, Atlanta GA 30332, United States}
\alsoaffiliation{Renewable and Sustainable Energy Institute, University of Colorado Boulder, Boulder CO 80303, United States}

\author{Christian Ahläng}
\affiliation{Physics, Faculty of Science and Engineering, \AA{}bo Akademi University, Porthansgatan 3, 20500 Turku, Finland}

\author{Stephen Barlow}
\affiliation{School of Chemistry \& Biochemistry, Georgia Institute of Technology, Atlanta GA 30332, United States}
\alsoaffiliation{Renewable and Sustainable Energy Institute, University of Colorado Boulder, Boulder CO 80303, United States}

\author{Mathias Nyman}
\affiliation{Physics, Faculty of Science and Engineering, \AA{}bo Akademi University, Porthansgatan 3, 20500 Turku, Finland}

\author{Seth R. Marder}
\affiliation{School of Chemistry \& Biochemistry, Georgia Institute of Technology, Atlanta GA 30332, United States}
\alsoaffiliation{Renewable and Sustainable Energy Institute, University of Colorado Boulder, Boulder CO 80303, United States}
\alsoaffiliation{Department of Chemical and Biological Engineering, University of Colorado Boulder, Boulder CO 80303, United States}
\alsoaffiliation{Department of Chemistry, University of Colorado Boulder, Boulder CO 80303, United States}
\alsoaffiliation{Chemistry and Nanoscience Center, National Renewable Energy Laboratory, Golden CO 80401, United States}

\author{Ronald Österbacka}
\affiliation{Physics, Faculty of Science and Engineering, \AA{}bo Akademi University, Porthansgatan 3, 20500 Turku, Finland}

\title{Cross-Linking of Doped Organic Semiconductor Interlayers for Organic Solar Cells: Potential and Challenges}

\begin{document}

\begin{abstract}
Solution-processable interlayers are an important building block for the commercialization of organic electronic devices such as organic solar cells. Here, the potential of cross-linking to provide an insoluble, stable and versatile charge transport layer based on soluble organic semiconductors is studied. For this purpose, a photo-reactive tris-azide cross-linker is synthesized. The capability of the small molecular cross-linker is illustrated by applying it to a p-doped polymer used as a hole transport layer in organic solar cells. High cross-linking efficiency and excellent charge extraction properties of the cross-linked doped hole transport layer are demonstrated. However, at high doping levels in the interlayer, the solar cell efficiency is found to deteriorate. Based on charge extraction measurements and numerical device simulations, it is shown that this is due to diffusion of dopants into the active layer of the solar cell. Thus, in the development of future cross-linker materials, care must be taken to ensure that they immobilize not only the host, but also the dopants.
\end{abstract}

\section{Introduction}
Doping is a key technology to tune the electronic properties of semiconductor devices. In inorganic semiconductors, doping is usually achieved by substituting impurity atoms into the crystal lattice. In organic materials, doping was first carried out using halogens and alkali metals as p- and n-dopants, respectively, leading, for example, to the discovery of conductive polymers in the late 1970s.\cite{Shirakawa1977} However, dopants forming small atomic ions soon proved to be unsuitable for practical applications due to the strong tendency of these ions to diffuse in organic host systems. Instead, molecular doping with larger dopant molecules has become standard practice in organic semiconductors.\cite{Lussem2013,Salzmann2016,Jacobs2017} Through electron transfer reactions, molecular dopants either oxidize or reduce the host material, which can increase the electrical conductivity by orders of magnitude. Molecularly doped materials are widely applied in organic electronic devices such as light-emitting diodes\cite{Pfeiffer2003} and thermoelectric generators.\cite{Liang2018,Untilova2020}

In organic solar cells~(OSCs), the active layer of which typically consists of a phase-separated network of electron donating and accepting materials, doping is much less common. Unintentional doping of the active layer is often considered detrimental for device performance due to undesired space charge effects,\cite{Dibb2013,Deledalle2015,Nyman2016,Sandberg2019} although it has been shown recently that a moderate doping level can have a positive effect on device performance in certain device configurations.\cite{Nyman2021} However, there has been an increased interest in doped \textit{interface} layers for OSCs in recent years. With the emergence of novel acceptor materials reducing voltage losses in the bulk,\cite{Hou2018,Armin2021} the quality of the contact interfaces has become of utmost importance. Ideally, the contacts should be Ohmic while avoiding losses through surface recombination, i.e., extraction of charge carriers at the ``wrong" contact.\cite{Sandberg2014,Sandberg2019b,Scheunemann2019} Such a situation is realized by sandwiching the active layer between two charge-selective interlayers that conduct only one type of charge carrier, namely holes at the anode (high work function electrode) and electrons at the cathode (low work function electrode). Doped organic interlayers are very attractive for this approach, since they facilitate not only precise control over the conductivity of electrons and holes, but also over the energy levels and Fermi level at the contact interface.\cite{Warren2019}

Methods that have been used to prepare doped organic thin films include co-evaporation,\cite{Walzer2007} soft-contact transfer lamination,\cite{Dai2014} and solid-state diffusion\cite{Kang2016}. However, from the commercialization point of view, there is a large interest in vacuum-free solution-based techniques that can be integrated in roll-to-roll processes. There are, in principle, two ways to process doped interlayers from solution: either the host and dopant are co-deposited from the same solution,\cite{Yim2008,Suh2019} or both are deposited sequentially and then let diffuse into each other.\cite{Scholes2015,Guillain2016,Kolesov2017} Regardless of which method is used, one challenge is that the processing of the subsequent layers must not redissolve the doped interlayer or change its electrical properties. The problem is usually addressed by using orthogonal solvents for the individual layers, but this limits the choice of host and dopant materials.

In this paper, we study the potential of cross-linking to provide stable and versatile doped organic interlayers for fully solution-processed organic electronic devices. Small molecule cross­-linkers have been widely used in literature to ``lock'' the morphology of the active layer in OSCs by establishing covalent bindings.\cite{Png2010,Wantz2014,Derue2014,Deb2015,Rumer2015,Rumer2015b,Kahle2017} Here, we adapt the concept to stabilize a p-doped polymer used as hole-transporting layer. Using a tris-azide cross-linker, we demonstrate that cross-linked interlayers can withstand the subsequent deposition of the active layer using the same solvent and provide similar performance in OSCs as the reference material PEDOT:PSS. We also examine the effect of the doping concentration and find that the device performance is maximized for ca.~\unit[4]{wt\%} of dopant. Combining transient charge extraction measurement and numerical simulations, we show that device performance at higher wt\% is restricted by diffusion of dopant molecular ions into the active layer. Our results provide a detailed understanding of the effect of doping of the active layer on the device performance.

\section{Results and Discussion}

\subsection{Materials and Cross-Linking Procedure}

\begin{figure*}[t]
\includegraphics[width=0.8\textwidth]{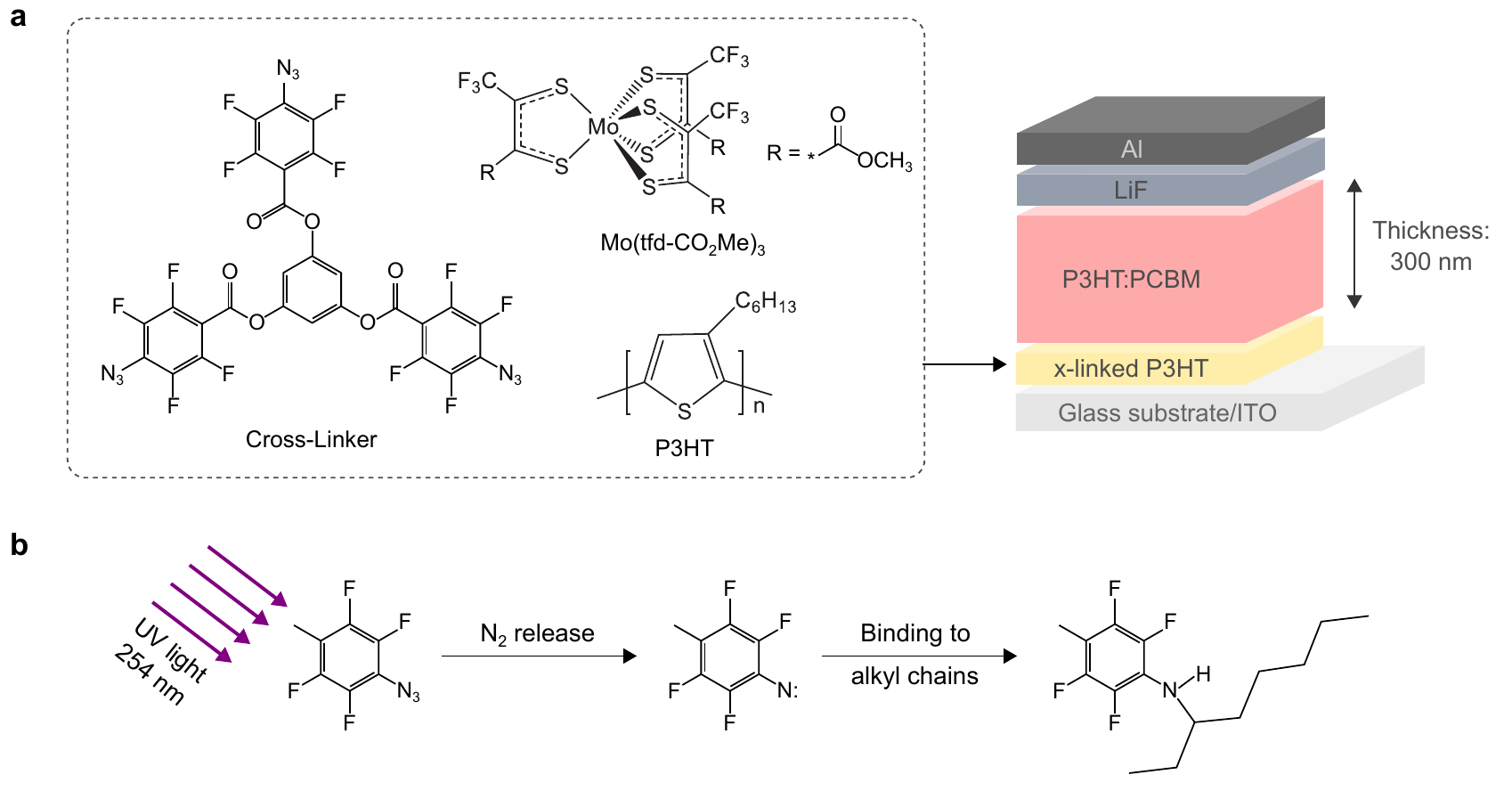}
\caption{(a)~Chemical structure of the materials used for the preparation of cross-linked~(x-linked) doped HTLs, the cross-linker~\crosslinker, the dopant~\dopant, and the polymeric host~P3HT. On the right side, the OSC~device structure used in this study is depicted. The devices are based on an active layer of P3HT:PCBM with exceptionally low recombination coefficients in the bulk, which allows the use of thick junctions of~\unit[300]{nm}. (b)~Schematic illustration of the UV-activated cross-linking mechanism.}
\label{fig:fig1}
\end{figure*}

Figure~\ref{fig:fig1}a illustrates the key materials and the device structure used in this study. As a model system for a doped hole transport layer~(HTL), we chose poly(3-hexylthiophene)~(P3HT) molecularly doped with molybdenum tris[1-(methoxycarbonyl)-2-(trifluoromethyl)-ethane-1,2-dithiolene]~(\dopant). Efficient hole collection was demonstrated for this particular system when co-deposited from solution and incorporated into bulk-heterojunction OSCs via lamination.\cite{Dai2014,Dai2015} Here, we study the potential of replacing the lamination process by cross-linking the interlayer, which would mark an important step towards fully solution-processable OSCs. For this purpose, a small molecule cross-linker~\crosslinker based on photo-reactive azide functional groups was synthesized as described in the Supporting Information. This compound has previously been reported as a precursor to fire-retardant materials, but has not been investigated as a cross-linker.\cite{Tang2016} Several previous approaches use this type of small molecule cross-linker, but often including only two azide groups.\cite{Png2010,Derue2014,Rumer2015b} The cross-linker used in this work comprises three azide groups instead of two to increase the cross-linking efficiency, although recently a molecule with four azides has been reported.\cite{Watson2017,Kim2020} In addition, in one recent study an azide has been directly attached to an n-dopant, allowing the dopant ion to be covalently anchored to a fullerene host material.\cite{Reiser2019}

The working principle of the cross-linker is illustrated in Figure~\ref{fig:fig1}b. Cross-linking is initialized by illumination with UV light~(wavelength \unit[254]{nm}), which leads to the release of \ce{N2} and the formation of a highly reactive nitrene. The nitrene then ideally reacts with the alkyl side chains of the polymer, in our case the hexyl groups of P3HT. The result is a network of covalently bound polymer chains that is insoluble to common solvents. High cross-linking efficiency was obtained with the new tris-azide molecule, as indicated by short UV illumination times of $<\unit[1]{min}$ and low cross-linker concentrations being required to afford insoluble films. The ratio of polymer monomers to cross-linker molecules was optimized experimentally by varying the cross-linker concentration in thin P3HT films and rinsing the UV-treated films with solvent by spin-coating. An insoluble film was achieved at a molar ratio as low as 100:1 P3HT:cross-linker, see Figure~S6 in the Supporting Information.

To clarify the effect of cross-linking and rinsing on the electrical properties, we carried out four-point probe conductivity measurements on thin films on glass. As a reference, we studied a \unit[2]{wt\%} doped P3HT film, resulting in a conductivity of~$\unit[3 \times 10^{-4}]{S\,cm^{-1}}$. Upon adding the cross-linker and UV-treating the film, the conductivity decreased by roughly a factor of three to~$\unit[9 \times 10^{-5}]{S\,cm^{-1}}$. A small decrease in conductivity can be expected upon adding the cross-linker, since it may induce changes in film morphology and possibly also affect the doping efficiency. After rinsing of the cross-linked and doped P3HT film with 1,2-dichlorbenzene, which is the solvent used for subsequent preparation of the active layer in the OSC devices described below, the conductivity dropped further by more than an order of magnitude to~$\unit[4 \times 10^{-6}]{S\,cm^{-1}}$. The significant decrease upon rinsing the film suggests that the dopants are not fully insolublized or spatially stable in the cross-linked polymer network. However, the insoluble cross-linked film still enables the fabrication of multilayer structures from solution and, even after rinsing, the film exhibits an adequate conductivity for use as an extraction layer~(\textit{vide infra}). The results from the conductivity measurements are summarized in Table~\ref{tab:cond}.

\begin{table}[t]
\caption{Conductivity of \unit[2]{wt\%}~doped P3HT films with and without cross-linking.}
\begin{tabular}{ll}
\toprule
Layer & Conductivity $\left[\unit{S\,cm^{-1}} \right]$ \\
\midrule
Doped P3HT & $3 \times 10^{-4}$  \\
Doped P3HT, x-linked & $9 \times 10^{-5}$ \\
Doped P3HT, x-linked and rinsed & $4 \times 10^{-6}$ \\
\bottomrule
\end{tabular}
\label{tab:cond}
\end{table}

\subsection{Photovoltaic Performance}
To test the functionality of the cross-linked doped P3HT films, we implemented them as HTL in OSCs with a standard device architecture~(see Figure~\ref{fig:fig1}a). For this purpose, the photoactive layer, consisting of a blend of P3HT as donor and the fullerene derivative phenyl-\ce{C61}-butyric acid methyl ester~(PCBM) as acceptor, was spin-coated directly on top of the cross-linked interlayer. Notably, the P3HT:PCBM blend layer was prepared in a manner that leads to an exceptionally low bimolecular recombination coefficient of $k_2 \sim \unit[10^{-19}]{m^3s^{-1}}$ in the bulk.\cite{Wilken2021} This enabled us to study devices with a relatively large absorber thickness of~\unit[300]{nm} without notable transport losses. Such thick layers are not only considered an important prerequisite for low-cost fabrication using printing techniques,\cite{Meredith2018,Chang2021} but also particularly susceptible to space charge effects and therefore ideally suited to investigate the impact of a possible diffusion of dopants from the HTL into the active layer.

\begin{figure*}[t]
\includegraphics[width=0.8\textwidth]{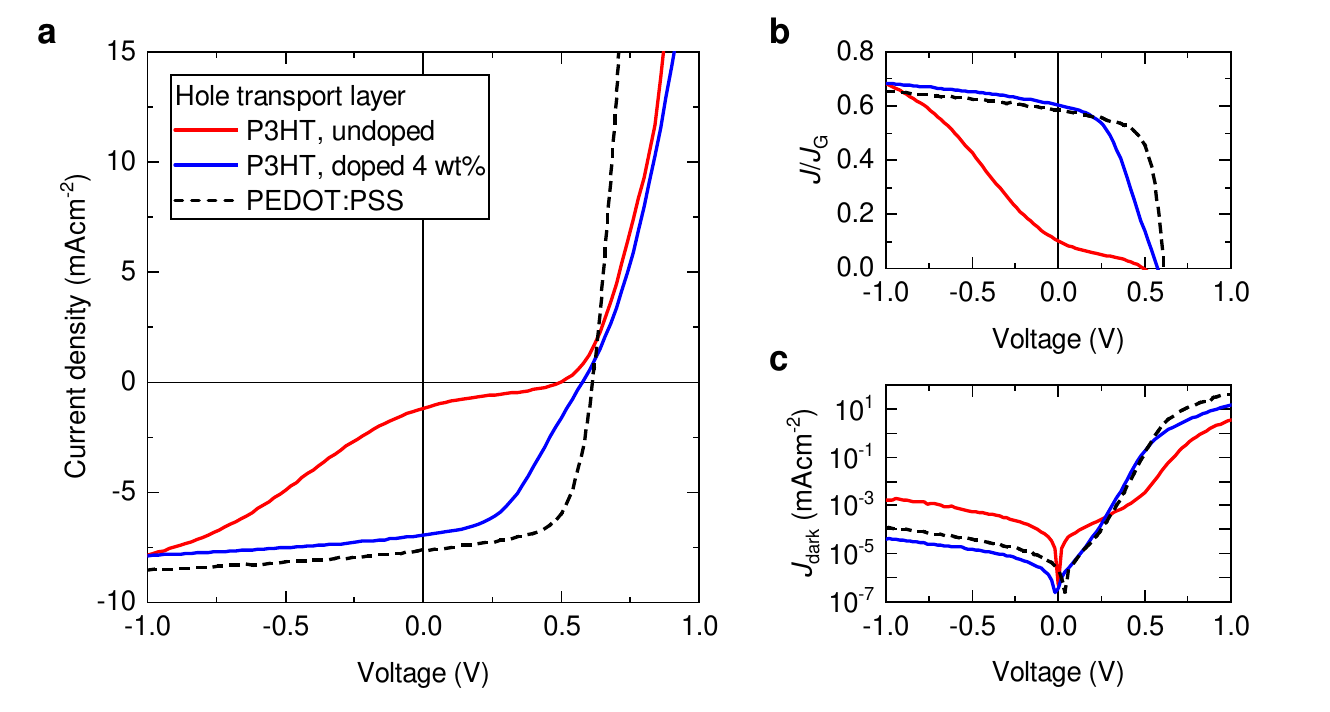}
\caption{(a)~Measured $J$--$V$ curves of P3HT:PCBM solar cells with a HTL of cross-linked undoped P3HT, cross-linked P3HT doped with \unit[4]{wt\%} \dopant, and the reference material PEDOT:PSS, respectively. (b)~Current density normalized to the generation current~$J_G$ calculated with a transfer-matrix optical model. The higher $\jsc$ in the PEDOT:PSS reference device can be explained by less parasitic absorption in the HTL. (c)~Current--voltage curves measured in the dark.}
\label{fig:fig2}
\end{figure*}

Figure~\ref{fig:fig2}a shows current--voltage~($J$--$V$) curves measured under illumination with simulated sunlight. As can be seen, very poor device performance is obtained when cross-linked P3HT is used as HTL without doping. In particular, the $J$--$V$~curve shows a pronounced S-shape, which is characteristic of a device limited by contact-related issues.\cite{Wagenpfahl2010,Sandberg2014,Wilken2014} This indicates that undoped P3HT does not provide a suitable hole-collecting interface, in agreement to literature.\cite{Dai2014} The device performance is significantly improved when using cross-linked P3HT doped with \dopant instead. Figure~\ref{fig:fig2}a shows this exemplarily for a doping level of \unit[4]{wt\%}, which has been identified as optimum concentration in previous reports.\cite{Dai2014,Dai2015} We also fabricated devices with an HTL of the very commonly used conducting polymer PEDOT:PSS as reference. While the photocurrent shows a very similar voltage dependence under reverse bias and around the short-circuit current~($\jsc$) for the PEDOT:PSS and the doped P3HT device, its absolute value is slightly higher for the reference device. We attribute the latter to more absorption of visible light by the P3HT than by PEDOT:PSS, rather than to electrical issues. This is evidenced by Figure~\ref{fig:fig2}b where we normalize the total current~$J$ to the generation current $J_G$, which we calculated using transfer-matrix optical modeling as detailed in the Supporting Information. The ratio~$J/J_G$ serves as a measure for the extraction efficiency and is not affected by parasitic absorption in the contacts and interfacial layers. As can be seen in Figure~\ref{fig:fig2}b, virtually identical extraction efficiencies are obtained around~$\jsc$ for the doped P3HT and the PEDOT:PSS reference device. We therefore conclude that the cross-linked  doped interlayers are in principle well suited for hole extraction in OSCs.

However, a different behavior to the PEDOT:PSS reference arises when the devices are operated closer to the open-circuit voltage~($\voc$). The doped P3HT device displays lower forward currents and also a slight S-shape around~$\voc$, which leads to a reduced fill factor compared to the reference device. This is further illustrated in Figure~\ref{fig:fig2}c, which shows $J$--$V$ curves measured in the dark. The main deviation between the doped P3HT and the PEDOT:PSS device occurs at high voltages in the forward direction. Therefore, it is mainly resistive losses~(i.e., large series resistance) that causes the reduction of the fill factor. Interestingly, the opposite trend is observed when analyzing the reverse part of the dark $J$--$V$ curves. Here it is the doped P3HT device that shows the superior performance in terms of a lower dark saturation current, that is, the diode leakage current under negative applied voltage. The dark saturation current is a key indicator of the dominant recombination mechanism.\cite{Cuevas2014,Tvingstedt2016,Sandberg2019b} Assuming that the recombination in the bulk of the active layer remains unchanged by the choice of HTL, we anticipate that doped P3HT provides an interface with reduced surface recombination losses compared to PEDOT:PSS.

\begin{figure*}[t]
\includegraphics[width=0.8\textwidth]{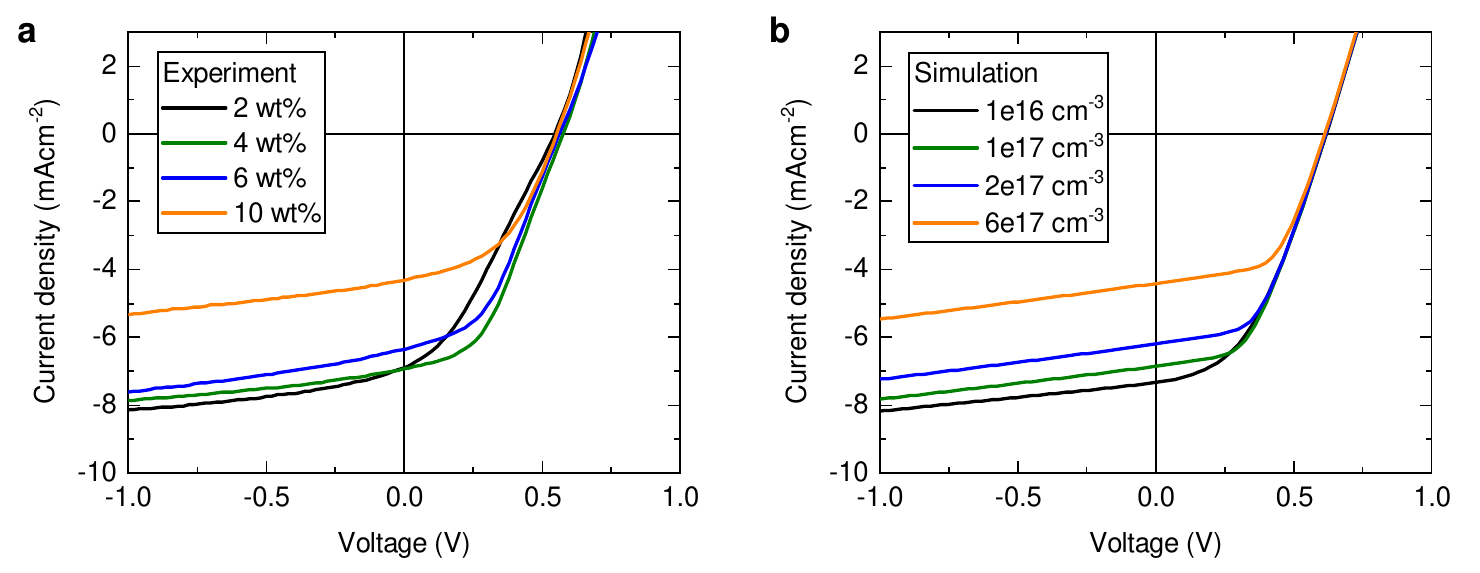}
\caption{(a)~Experimental $J$--$V$ curves for devices with different doping levels in the cross-linked P3HT interlayer. (b)~Results of drift--diffusion simulations assuming different bulk doping concentrations $N_p$ in the active layer. In the model, a metal--insulator--metal approach was chosen and the properties of the contacts were held constant. The experimentally observed variation of $\jsc$ can be explained solely by undesirable doping of the active layer through diffusion of dopants from the HTL.}
\label{fig:fig3}
\end{figure*}

To test whether the resistive properties of the cross-linked P3HT can be improved, we performed a doping concentration study. In principle, increasing the doping is expected to lead to a higher density of mobile holes in the HTL and thereby a higher conductivity. Figure~\ref{fig:fig3}a shows $J$--$V$ curves for devices with cross-linked P3HT doped with 2, 4, 6 and \unit[10]{wt\%} \dopant, respectively. It can be seen that the fill factor is indeed improved by increasing the doping level from 2 to \unit[4]{wt\%}, in agreement with previous work on laminated doped HTLs.\cite{Dai2014} However, further increasing the doping level in the P3HT~interlayer degrades the device performance by reducing the photocurrent. This is particularly pronounced for the device with \unit[10]{wt\%} doped P3HT, where~$\jsc$ drops significantly from $6.9$ to $\unit[4.3]{mA\,cm^{-2}}$ compared to the optimum doping of \unit[4]{wt\%}. A possible explanation for the reduced photocurrent at high doping levels is diffusion of dopants and/or dopant ions into the active layer, which would then give rise to undesired bulk doping.\cite{Dibb2013,Deledalle2015,Nyman2016,Sandberg2019}

\subsection{Dopant Diffusion Degrades Device Performance}
In order to clarify whether diffusion of dopants into the active layer is taking place, we measured the doping concentration in the active layer of complete OSC devices using the charge extraction by linearly increasing voltage technique in the doping-induced capacitive regime~(doping-CELIV).\cite{Sandberg2014b} In the doping-CELIV method, capacitance--voltage data from the measured capacitive extraction currents are analyzed using Mott--Schottky theory. The measured doping concentration can be determined as a function of the distance from one of the electrodes in the solar cell, i.e., a depth profile of any dopant migration into the active layer can be obtained.\cite{Nyman2016,Nyman2017} Current transients for an undoped device in the dark consist of a constant current plateau given by the displacement current~$j_0$ from the geometric capacitance of the device,
\begin{equation}
j_0 = \varepsilon \varepsilon_0 \frac{A}{L},
\end{equation}
where $\varepsilon$ is the relative dielectric constant, $\varepsilon_0$ the vacuum permittivity, $A$ the voltage rise speed and $L$ the active-layer thickness. In the case of sufficiently high doping, an additional time-dependent extraction current~$\Delta j(t)$ caused by the extraction of doping-induced charge carriers is added to the transient current response, $j(t) = j_0 + \Delta j(t)$. Figure~S8 in the Supporting Information shows that both the reference device with a PEDOT:PSS interlayer and the device with an undoped P3HT interlayer show a constant current plateau given by~$j_0$. This means that the active layer is undoped for these devices.

\begin{figure*}[t]
\includegraphics[width=0.8\textwidth]{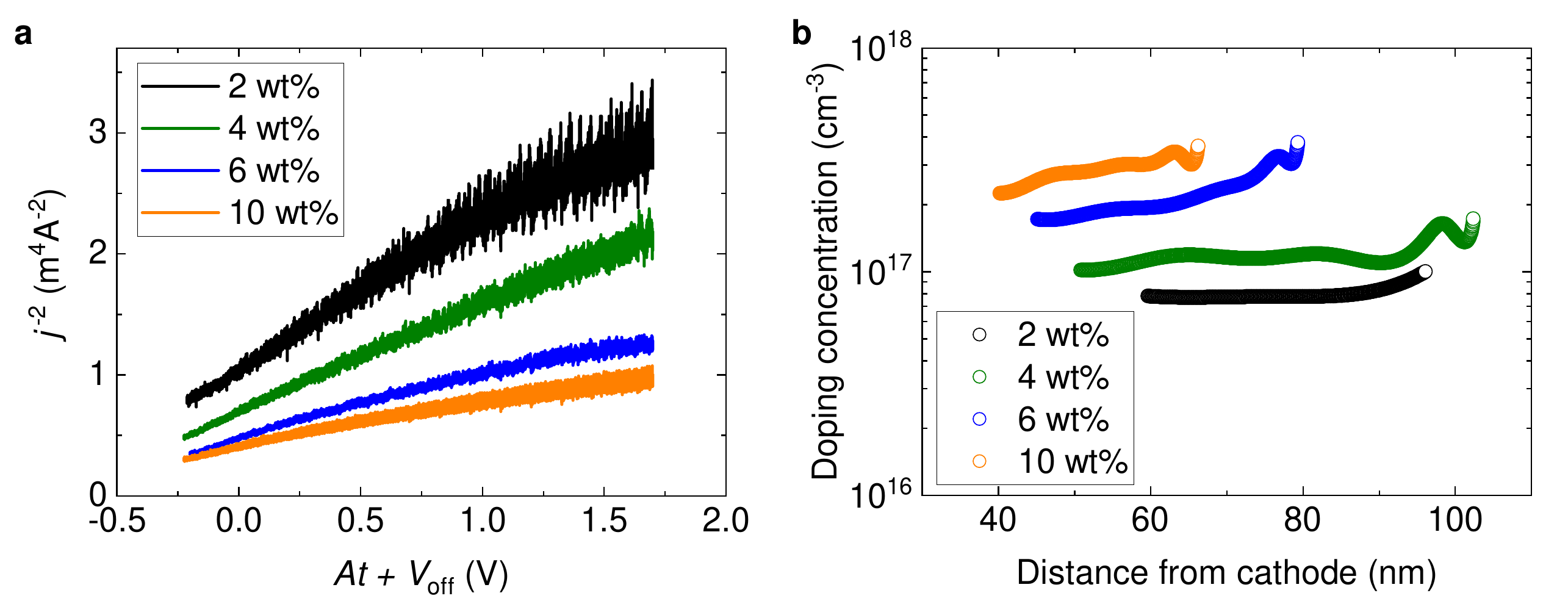}
\caption{Result of doping-CELIV measurements to determine the doping concentration in the active layer. (a)~Inverse square of the measured extraction current in the capacitive regime plotted as a function of the applied voltage given by $At + V_\text{off}$, where $V_\text{off}$ is a steady-state offset voltage. (b)~Doping concentration as a function of the distance from the cathode~(i.e., the LiF/Al top electrode).}
\label{fig:fig4}
\end{figure*}

For devices with a doped P3HT layer at the anode, the current transients have a different shape due to an additional time-dependent current~$\Delta j(t)$ originating from doping-induced free charge carriers in the active layer. In Figure~\ref{fig:fig4}a, the doping-induced capacitive regime of the current transient data is plotted in a Mott--Schottky representation, i.e., the inverse square of the extracted capacitive current as a function of the applied voltage. The doping concentration can be determined from the slope of the plot, and Figure \ref{fig:fig4}b shows the resulting depth profile~(for details on the doping-CELIV analysis, see experimental section). Even for the lowest doping level of~\unit[2]{wt\%} in the cross-linked interlayer, the doping concentration in the active layer is close to~$\unit[10^{17}]{cm^{-3}}$. With increasing doping in the HTL, the doping concentration in the active layer also increases up to almost~$\unit[3 \times 10^{17}]{cm^{-3}}$ for a doping level of~\unit[10]{wt\%}. The doping concentration slightly increases when moving away from the cathode towards the doped interlayer at the anode, from which the dopants and doping-induced carriers originate from. It should be noted that the depth profiles are ultimately restricted by the magnitude of the voltage that can be applied without risking dielectric breakdown. Further restriction arises from significant leakage currents through the solar cell diode at higher voltages. The leakage current can be corrected for to some extent, but at high voltages there will be a larger error in the analysis, corresponding to the data at longer distances from the cathode in Figure~\ref{fig:fig4}b. Due to this restriction, only 30 to~\unit[50]{nm} of the \unit[300]{nm} thick active layer could be reliably measured for the devices under test. 

In order to understand the effect of the apparent bulk doping on the device performance, we performed numerical device simulations using a one-dimensional drift--diffusion model.\cite{Burgelman2000,Scheunemann2019,Wilken2020} Since we are only interested in properties of the active layer at this point and to keep the number of input parameters to a minimum, we chose a metal--insulator--metal approach, i.e., the bulk-heterojunction blend was modeled as an effective semiconductor sandwiched between two charge-selective contacts. Table~\ref{tab:params} lists the key input parameters of the model. The values related to the P3HT:PCBM blend layer were taken from a recent study\cite{Wilken2021} in which the same processing protocol was used as in this work. The non-uniform generation rate profile of the thick-film devices was fully taken into account by coupling the drift--diffusion simulator with the transfer-matrix model. In the first step, we modeled the PEDOT:PSS reference device assuming the active layer to be undoped. Figure~S9 in the Supporting Information shows an excellent agreement between simulated and measured $J$--$V$ curves, which justifies the use of the parameter set in Table~\ref{tab:params} for the given system.

\begin{table}
\caption{Input parameters for the drift--diffusion simulations.}
\begin{tabular}{lll}
\toprule
Parameter & Symbol & Value\\
\midrule
Thickness & $L$ & \unit[300]{nm} \\
Effective band gap & $E_g$ & \unit[1.0]{eV}\\
Injection barrier height & $\varphi$ & \unit[0.1]{eV}\\
Relative permittivity & $\varepsilon$ & 3.5\\
Effective density of states & $N_C$, $N_V$ & $\unit[10^{20}]{cm^{-3}}$\\
Electron mobility & $\mu_n$ & $\unit[8.9 \times 10^{-4}]{cm^{2}V^{-1}s^{-1}}$\\
Hole mobility & $\mu_p$ & $\unit[1.3 \times 10^{-4}]{cm^{2}V^{-1}s^{-1}}$\\
Recombination coefficient & $k_2$ & $\unit[1 \times 10^{-13}]{cm^{3}s^{-1}}$\\
Doping density (p-type) & $N_p$ & variable\\
\bottomrule
\end{tabular}
\label{tab:params}
\end{table}

We then assumed p-doping in the numerical model by introducing acceptors of variable concentration~$N_p$ into the active layer. For simplicity, we assumed the dopants to be  homogeneously distributed over the thickness, which is at least for parts of the active layer justified by the doping-CELIV profiles. Figure~\ref{fig:fig3}b shows that by simply increasing~$N_p$, the numerical model reproduces the experimental $J$--$V$ curves reasonably well. In particular, the strong reduction of~$\jsc$ for the \unit[10]{wt\%} device, while the fill factor remains about the same, is well captured by the model. A similar behavior for thick-film OSCs with an unintentionally doped active layer was reported by Deledalle~et~al.\cite{Deledalle2015} It should be mentioned that in particular for high doping levels in the HTL, the values of $N_p$ that had to be assumed in the numerical model to quantitatively describe the measured photocurrents are slightly higher than suggested from the CELIV measurements~(see Supporting Information, Figure~S10). This indicates that the actual doping profiles are more complex than the homogeneous distribution we assumed in the model, and an increase of~$N_p$  towards the anode can be expected. However, the qualitative statement that the degradation of the device performance can be explained by undesired doping in the bulk alone remains unaffected.

\begin{figure*}[t]
\includegraphics[width=\textwidth]{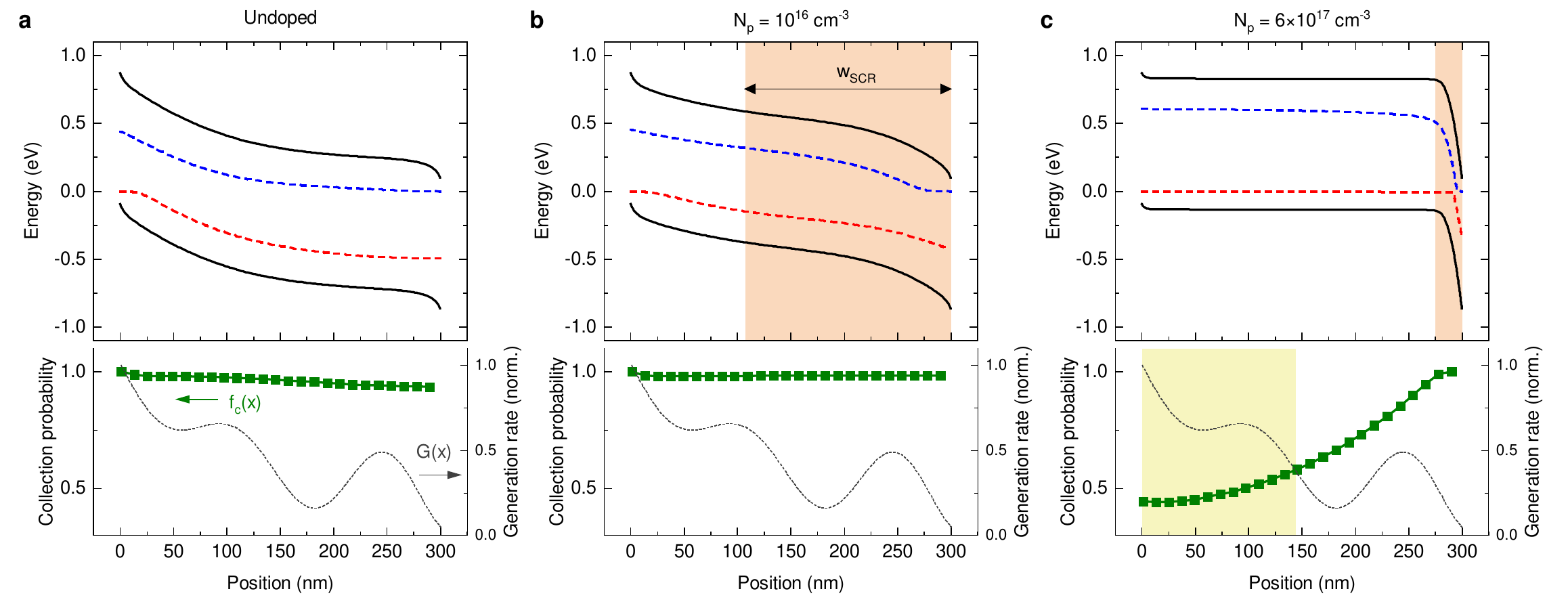}
\caption{Simulated band diagrams under short-circuit conditions and corresponding charge collection probability~$f_c(x)$ for a device with an (a)~undoped, (b)~medium doped and (c)~highly doped active layer. The orange shaded areas in the band diagrams indicate the width of the doping-induced space charge region according to Equation~(\ref{eq:wscr}). For the highly doped device, the region of low $f_c(x)$ values conincides with the highest photo-generation~(yellow shaded area), with explains the significant loss in photocurrent.}
\label{fig:fig5}
\end{figure*}

Figure~\ref{fig:fig5} illustrates the effect of bulk doping with simulated energy band diagrams under short-circuit conditions. Also shown is the collection probability~$f_c(x)$, that is, the probability that a charge carrier generated at position $x$ reaches the respective contact and is extracted there.\cite{Wilken2020,Kirchartz2012,Kirchartz2016} Based on the doping level, two limiting cases can be distinguished for the P3HT:PCBM devices under consideration. For no or very low doping~(Figure~\ref{fig:fig5}a), the energy bands are determined by the given mobility contrast~($\mu_n/\mu_p \approx 7$) in the active layer. Since holes move much more slowly than electrons, a space charge region~(SCR) is formed near the anode, while the electric field is screened in the vicinity of the cathode. However, due to the greatly reduced recombination rate and the resulting long diffusion length, electrons and holes are also effectively collected from the quasi-neutral region in which $f_c$ is only slightly smaller than in the SCR~(where $f_c \rightarrow 1$). Thus, we cannot make the common simplification of assuming of a step-like collection probability with 100\% collection in the SCR and 0\% collection in the field-free region.\cite{Dibb2013,Scheunemann2015} The other limiting case is a high doping concentration, exemplified for $N_p = \unit[6 \times 10^{17}]{cm^{-3}}$ in Figure~\ref{fig:fig5}c. In this case, a narrow SCR of only $\sim$\unit[25]{nm} is formed at the cathode, while the remaining $\sim$\unit[275]{nm} of the absorber remain field-free. The thickness~$w_\text{SCR}$ of the SCR agrees well with the theoretical expectation
\begin{equation}
w_\text{SCR} = \sqrt{\frac{2\varepsilon\varepsilon_0 (V_0 - V - kT/q)}{qN_p}},
\label{eq:wscr}
\end{equation}
where $V_0$ is the built-in voltage.\cite{Sandberg2019,Kirchartz2012} For high doping concentrations, the field-free region is so large that it exceeds the diffusion length, with the result that a significant fraction of the charge carriers recombine instead of being collected. Since the devices are illuminated from the anode side, a large fraction of the carriers is generated in a region with low~$f_c$, which explains the photocurrent losses in the \unit[10]{wt\%} device with the highest bulk doping according to the CELIV measurements. For a medium doping level, exemplified for $N_p = \unit[10^{16}]{cm^{-3}}$ in Figure~\ref{fig:fig5}b, the effects of doping and imbalanced charge transport compensate each other and the band diagrams are relatively homogeneous. The result is a nearly unity collection efficiency throughout the whole active layer. Hence, for the P3HT:PCBM blend system under consideration, a certain degree of bulk doping is even beneficial for the device performance. This provides a reasonable explanation of why the photocurrent~(corrected for parasitic absorption) in the optimized device with a \unit[4]{wt\%} doped~HTL is even slightly higher than in the PEDOT:PSS reference device~(see Figure~\ref{fig:fig2}b).

We can thus conclude from this section that while the cross-linked P3HT host withstands the following deposition steps, at least some of the dopant \dopant is released from the HTL and diffuses into the active layer. The higher the weight fraction of the dopant in the HTL, the more will diffuse into the active layer and cause bulk doping. A certain bulk doping level will not degrade the device performance and, in the case of the P3HT:PCBM blend investigated herein, can even compensate for the space charge effects due to imbalanced charge transport. However, if the bulk doping concentration is too high, the device photocurrent will drop, since a significant fraction of the charge carriers is generated in a field-free region with low collection probability.

\section{Conclusions}
In summary, we have investigated the potential of cross-linked doped organic semiconductors as hole transport layer in organic solar cells. As a model system, we studied interlayers of P3HT doped with \dopant that were cross-linked with a photo-reactive tris-azide cross-linker. We have shown that the extraction properties of the cross-linked doped P3HT are comparable to the reference material PEDOT:PSS. The fill factor, however, is found to be lower than in the reference devices, which is due to resistive losses close to the open-circuit voltage. This could not be avoided by increasing the doping in the hole transport layer, since a proportion of the dopants is found to diffuse into the active layer. We were able to detect the undesired bulk doping by CELIV measurements and to understand its effect using numerical simulations. At high doping concentrations, thick-film devices suffer significant losses in photocurrent since a large part of the charge carriers is generated in a field-free region where recombination losses occur. To avoid this, it is crucial to develop cross-linkers that not only stabilize the host material, but also the dopant in the doped interlayer.

\section{Experimental Section}

\paragraph{Materials} The cross-linker~\crosslinker was synthesized as described in the Supporting Information. The dopant~\dopant was synthesized as reported elsewhere.\cite{Dai2014} P3HT~(regioregularity $> 90\%$) and PCBM were purchased from Sigma-Aldrich. Poly(3,4-ethylene\-di\-oxy\-thio\-phene) polystyrene sulfonate~(PEDOT:PSS) was purchased from Ossila~(AI~4083).

\paragraph{Preparation of Cross-Linked HTLs} P3HT was mixed with the cross-linker in chlorobenzene at a molar ratio of 100:1~P3HT monomers:cross-linker molecules. The dopant was added at different weight percentages ranging from~2 to~\unit[10]{wt\%}. The concentration of P3HT in the solution was kept at~$\unit[5]{mg\,mL^{-1}}$ for all dopant ratios. The solution was filtered through a $\unit[0.2]{\mu m}$ PTFE filter. The P3HT films were spin-coated and annealed at~\unit[120]{$^{\circ}$C} for~\unit[5]{min} directly after the deposition. The annealed films were exposed to UV~light at a wavelength of~\unit[254]{nm} for~\unit[1]{min} in order to activate and complete the cross-linking mechanism. An UV~lamp with two \unit[8]{W} light tubes was used and samples were kept at a distance of roughly~\unit[10]{cm} from the light source during exposure. The films were afterwards rinsed with chlorobenzene to remove any residual soluble material.

\paragraph{Device Fabrication} Organic solar cells were prepared with the device structure glass\slash{}indium tin oxide (ITO)\slash{}HTL\slash{}P3HT:PCBM\slash{}LiF\slash{}Al. The ITO  substrates~(Präzisions Glas \& Optik GmbH) were structured by etching with~\ce{HCl}~(37--38\%). After ultrasonication in de-ionized water, acetone and isopropanol, the etched substrates were treated in a plasma cleaner. Subsequently, the HTL was deposited via spin-coating. Cross-linked~HTLs were spin-coated inside a nitrogen-filled glovebox. For the reference devices, the PEDOT:PSS~solution was filtered through a 0.45 $\unit{\mu}$m PVDF filter and deposited in ambient air. The devices were then transferred into the glovebox and dried on a hotplate for \unit[10]{min} at~\unit[120]{$^{\circ}$C}. For the active layer, P3HT and PCBM were mixed at 1:1 weight ratio and dissolved in 1,2-dichlorobenzene with a total concentration of~$\unit[45]{mg \,mL^{-1}}$. The active layer was deposited by static spin-coating at~\unit[1000]{min$^{-1}$} for~\unit[2:45]{min}. After deposition, the films were annealed at~\unit[150]{$^{\circ}$C} for~\unit[10]{min} inside a glass petri dish. The top contact, consisting of~\unit[0.8]{nm} LiF and~\unit[60]{nm} Al, was thermally evaporated through a shadow mask. The device area of about~\unit[4]{mm$^2$} was given by the overlap between the bottom and front electrode.

\paragraph{Measurements} Conductivity measurements were performed under nitrogen atmosphere using a four-point probe setup in a linear configuration. Films were prepared on~$\unit[2.5 \times 2.5]{cm^2}$ square shaped insulating glass slides for the measurements. Spring-loaded gold probes were used in the setup with a spacing of~\unit[1.79]{mm}. Measurements were performed using a source meter~(Keithley~2400) applying a current on the outer probes, while the voltage drop over the inner probes was measured. The measured voltage as a function of current showed a linear behavior in the measured regime. Conductivities were calculated using finite-size corrections.\cite{Smits1958} Current--voltage measurements of solar cell devices were carried out in ambient atmosphere using a source meter~(Keithley 2636) and an AM1.5 solar simulator~(Newport 92250A). CELIV~measurements were carried out with the sample mounted in a vacuum cryostat. A pulse generator~(SRS DG 535) and a function generator~(SRS DS 345) were used for generating the linearly increasing voltage pulse and an oscilloscope~(Keysight InfiniiVision DSOX3104T) was used to measure the corresponding current response. The voltage pulse was applied in reverse bias of the solar cell diodes in order to avoid injection of charge carriers and to extract charge carriers present in the active layer. Note that the voltage applied in reverse bias is by convention defined as positive in the CELIV~measurements. The pulse length was varied to ensure that the measurements are performed in the doping-induced capacitive regime.\cite{Sandberg2014b} This regime is reached at long enough pulse lengths where the transient is determined by the width of the space charge region caused by doping, $j(t) = \varepsilon\varepsilon_0 A w_\text{SCR}^{-1}$, where  $w_\text{SCR}$ is given by Equation~(\ref{eq:wscr}). The doping profile can then be obtained using
\begin{equation}
N_p(x) = \frac{2}{q\varepsilon\varepsilon_0 A^2} \left[{\frac{\diffd}{\diffd V}j(V)^{-2}}\right]^{-1},
\label{eq:Np}
\end{equation}
as reported previously.\cite{Nyman2016,Nyman2017}

\section{Acknowledgements}
This project has received funding from the Jane and Aatos Erkko foundation through the ASPIRE project and the European Union's Horizon 2020 research and innovation programme under the Marie Sk\l{}odowska-Curie grant agreement No~799801~(`ReMorphOPV'). Work at Georgia Tech was supported by the Air Force Office of Scientific Research~(FA9550-18-1-0499) and the Office of Naval Research~(N00014-21-1-2087 and N00014-21-1-2180). SD would like to thank the group members of Prof. Seth Marder's research group for their support during his research visit at Georgia Institute of Technology in the fall of 2017 and acknowledges funding from Vilho, Yrjö, and Kalle Väisälä foundation, Swedish Academy of Engineering Sciences in Finland and Otto A. Malm foundation. MN acknowledges funding from the Academy of Finland through project \# 326000.

\bibliography{ms}

\end{document}


\pagebreak

\section{Synthesis of the Cross-Linker}

\begin{figure}
\includegraphics[width=0.8\textwidth]{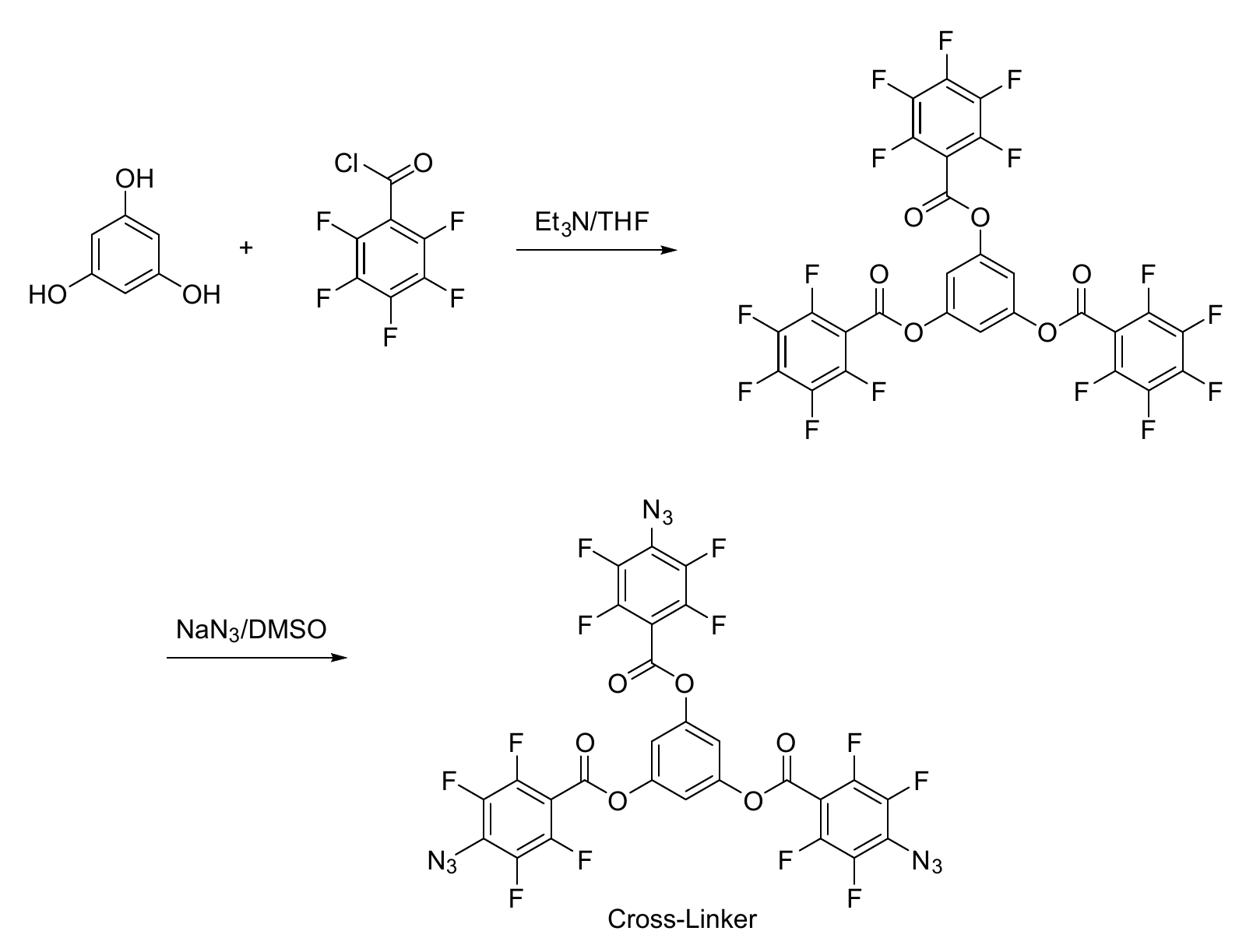}
\caption{Synthesis of the cross-linker used in this work.}
\label{fig:figS1}
\end{figure}

The crosslinker has previously been obtained from the reaction of 4-azido-2,3,5,6-tetra\-fluoro\-benzoyl chloride and benzene-1,3,5-triol.\cite{Tang2016} Here we used a different route, as shown Figure~\ref{fig:figS1}, the intermediate benzene-1,3,5-triyl tris(2,3,4,5,6-penta\-fluoro\-benzoate) being synthesized in a similar way to a previous report.\cite{Demyanov1991}

\paragraph{Benzene-1,3,5-triyl tris(2,3,4,5,6-penta\-fluoro\-benzoate).} A solution of benzene-1,3,5-triol~(\unit[1.8]{g}, \unit[14.3]{mmol}) in THF~(\unit[25]{mL}) and \ce{Et3N}~(\unit[4]{mL}) was added over~\unit[30]{min} to a solution of 2,3,4,5,6-penta\-fluoro\-benzoyl chloride (\unit[10.0]{g}, \unit[43.4]{mmol}) in THF~(\unit[25]{mL}) at room temperature under nitrogen. The reaction mixture was stirred at room temperature for~\unit[5]{h}. THF was removed under reduced pressure and water~(\unit[100]{mL}) was added, resulting in the appearance of a pink solid, which was collected by filtration, and purified through column chromatography on silica gel using di\-chloro\-methane/hexane~(1:1) as the eluent. After removal of solvents and drying under vacuum, the pure product was obtained as white solid~(\unit[5.2]{g}, 52\%). \ce{^1H} NMR (\unit[400]{MHz}, \ce{CDCl3}) $\delta$:~7.29 (s, 3H) ppm. \ce{^19F} NMR (\unit[376.5]{MHz}, \ce{CDCl3}) $\delta$:~-136.37 (m, 2F), -145.67 (m, 1F), -159.33 (m, 2F) ppm.

\paragraph{Benzene-1,3,5-triyl tris(4-azido-2,3,5,6-tetra\-fluoro\-benzoate) (Cross-Linker).} A solution of~\ce{NaN3} in a mixture of~DMSO and~\ce{H2O} (5:1~v/v, \unit[6.0]{mL}) was added over~\unit[6]{h} to a solution of benzene-1,3,5-triyl tris(2,3,4,5,6-penta\-fluoro\-benzoate)~(\unit[0.50]{g}, \unit[0.71]{mmol}) in a mixture of DMSO and THF~(1:1~v/v, \unit[10]{mL}) at room temperature. The reaction mixture was stirred at room temperature for~\unit[22]{h}; water~(\unit[100]{mL}) was then added; the resulting pink solid was collected by filtration and washed with methanol. The product was purified through column chromatography on silica gel using di\-chloro\-methane/hexane~(3:2) as the eluent. After removal of solvents under reduced pressure and drying under vacuum, the pure product was obtained as a white solid~(\unit[0.40]{g}, 73\%). \ce{^1H} NMR (\unit[400]{MHz}, \ce{CDCl3}) $\delta$:~7.26 (s, 3H) ppm. \ce{^19F} NMR (\unit[376.5]{MHz}, \ce{CDCl3}) $\delta$:~-137.00 (m, 2F), -150.23 (m, 2F) ppm.

\begin{figure}
\includegraphics[width=\textwidth]{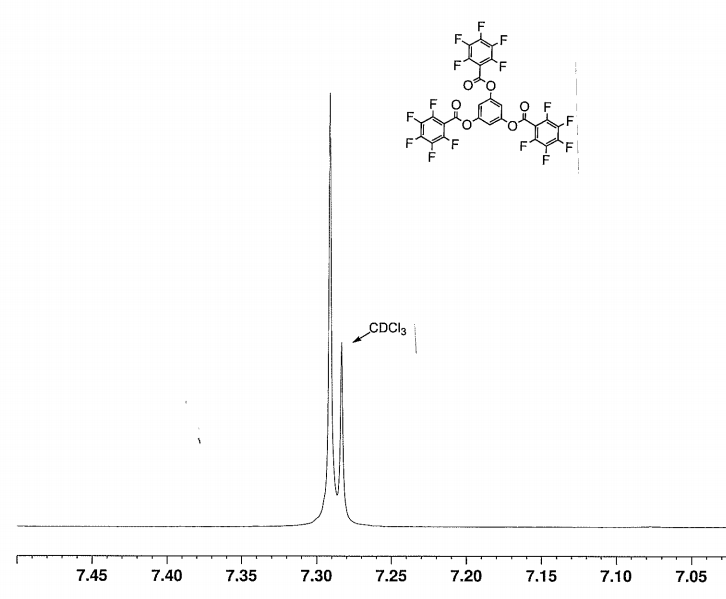}
\caption{\ce{^1H} spectrum for benzene-1,3,5-triyl tris(2,3,4,5,6-penta\-fluoro\-benzoate).}
\label{fig:figS2}
\end{figure}

\pagebreak

\begin{figure}
\includegraphics[width=\textwidth]{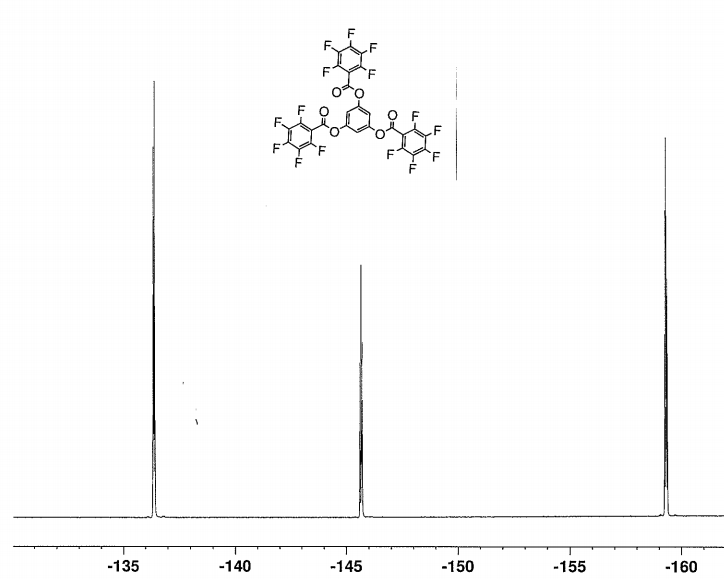}
\caption{\ce{^19F} spectrum for benzene-1,3,5-triyl tris(2,3,4,5,6-penta\-fluoro\-benzoate).}
\label{fig:figS3}
\end{figure}

\pagebreak

\begin{figure}
\includegraphics[width=\textwidth]{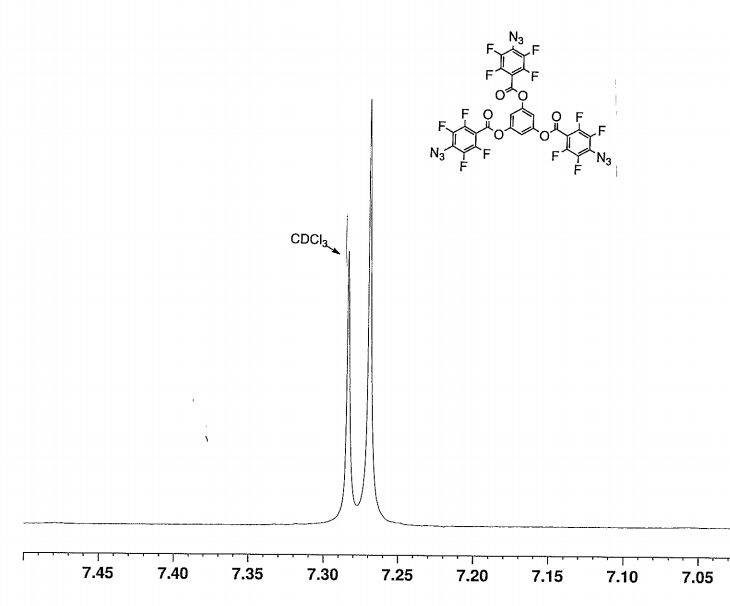}
\caption{\ce{^1H} spectrum for benzene-1,3,5-triyl tris(4-azido-2,3,5,6-tetra\-fluoro\-benzoate).}
\label{fig:figS4}
\end{figure}

\pagebreak

\begin{figure}
\includegraphics[width=\textwidth]{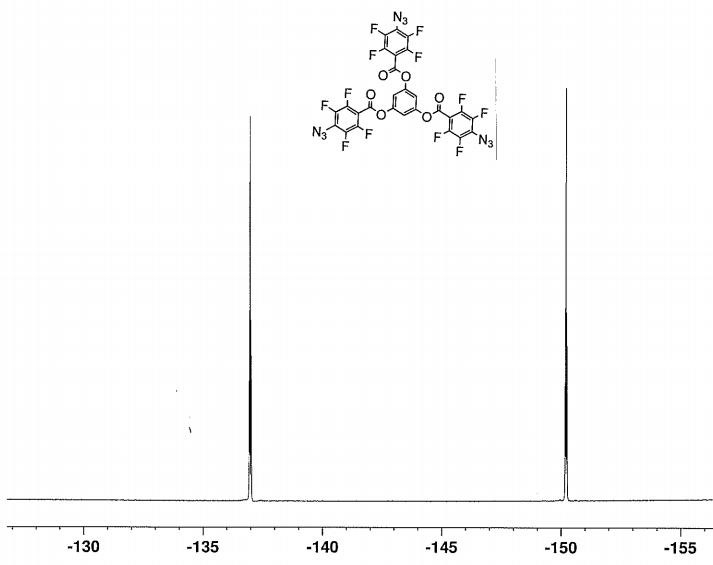}
\caption{\ce{^19F} spectrum for benzene-1,3,5-triyl tris(4-azido-2,3,5,6-tetra\-fluoro\-benzoate).}
\label{fig:figS5}
\end{figure}

\pagebreak

\begin{figure}
\includegraphics[width=\textwidth]{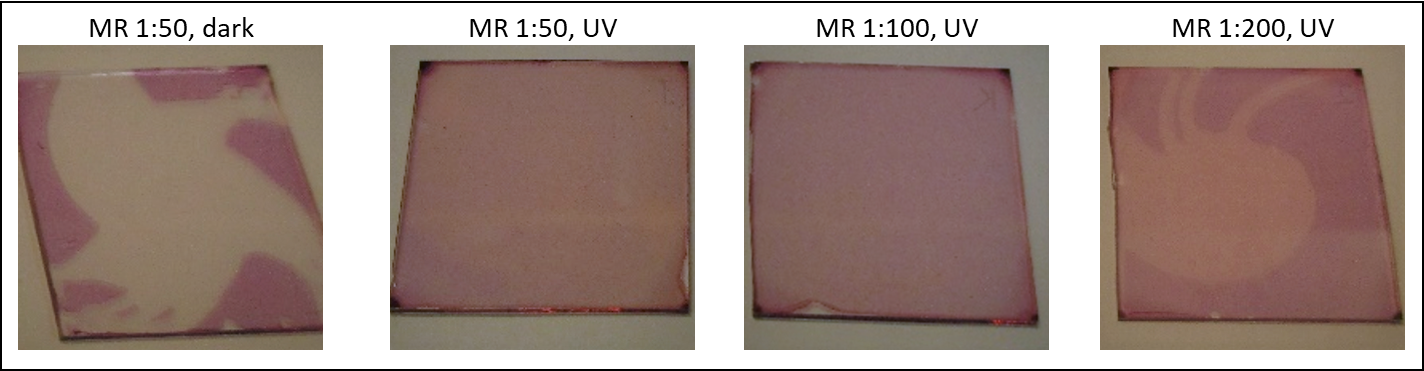}
\caption{Photographs of thin films of P3HT with the cross-linker added in different molar ratio~(MR) after rinsing with 1,2-dichlorobenzene. The left photo shows a film kept in the dark, while the films in the other three photos were treated with UV light to initialize the cross-linking mechanism.}
\label{fig:figS6}
\end{figure}

\pagebreak

\section{Transfer-Matrix Model}
\begin{figure}
\includegraphics[width=0.7\textwidth]{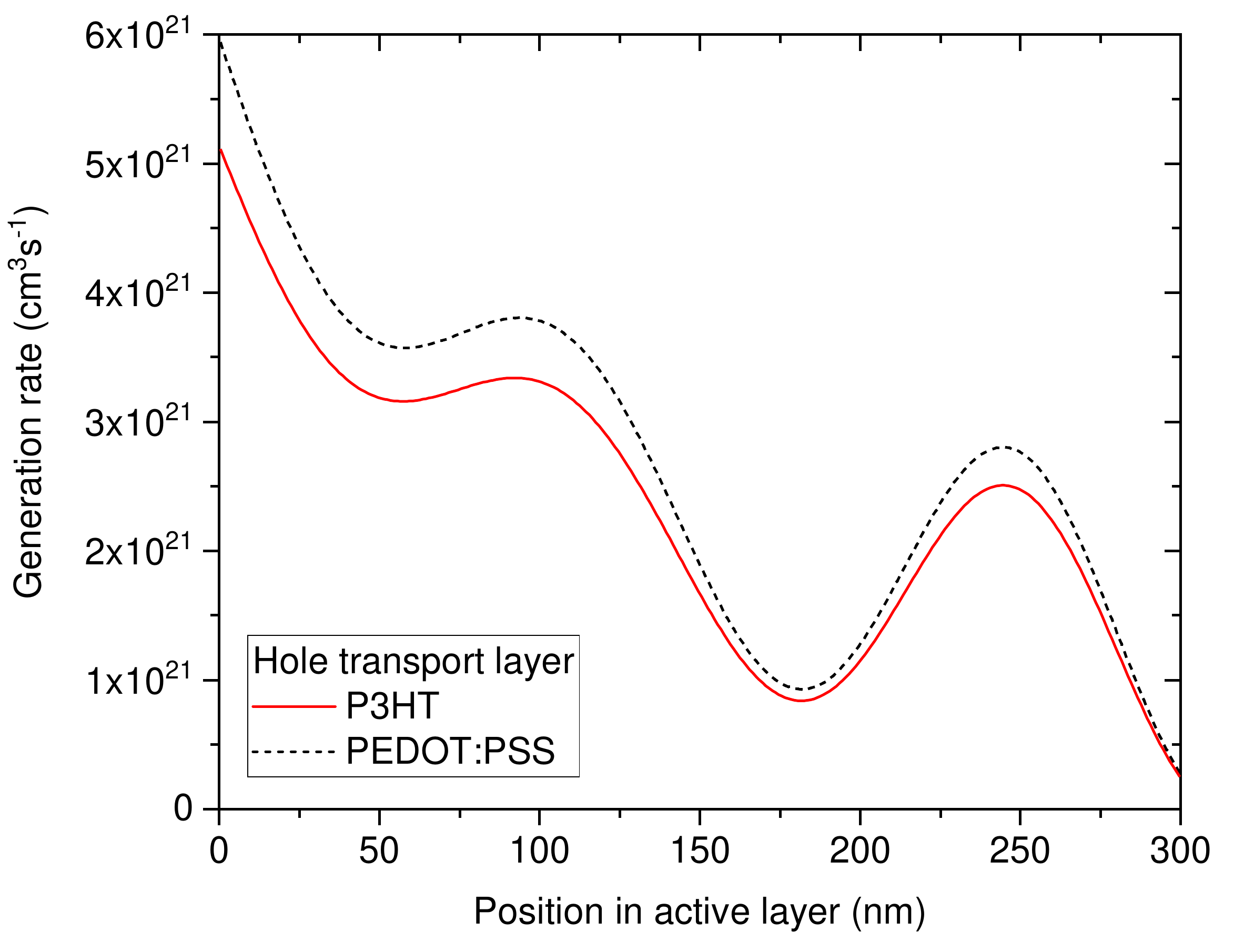}
\caption{Spatial generation profile~$G(x)$ for a P3HT:PCBM solar cell with a hole transport layer of doped P3HT and PEDOT:PSS, respectively.}
\label{fig:figS7}
\end{figure}

To calculate the spatial generation rate profile~$G(x)$ in the active layer, optical transfer-matrix simulations were performed using a MATLAB code.\cite{Burkhard2010} The optical constants of all involved material layers were taken from spectroscopic ellipsometry measurements published in previous reports.\cite{Wilken2015,Scheunemann2015,Wilken2020} Here, we made the assumption that the introduction of the dopant and the cross-linker has only negligible effect on the optical properties of the P3HT used as HTL, which is reasonable given the low weight percentage of the impurities. Figure~\ref{fig:figS7} shows the generation rate profile for a device with a HTL of P3HT and PEDOT:PSS, respectively. While the type of HTL has virtually no effect on the shape of~$G(x)$, its magnitude is significantly higher for the PEDOT:PSS device. The reason is that~PEDOT:PSS absorbs less light in the visible spectrum, i.e., shows a higher transparency. The generation profiles were used to calculate the generation current~$J_G$, that is, the maximum achievable photocurrent if there were no conversion and collection losses, via
\begin{equation}
J_G = q \int_0^L G(x) \diffd x,
\end{equation}
where $q$ is the elementary charge and $L$ is the active-layer thickness.

\pagebreak

\pagebreak

\section{CELIV Measurements}
\begin{figure}
\includegraphics[width=0.8\textwidth]{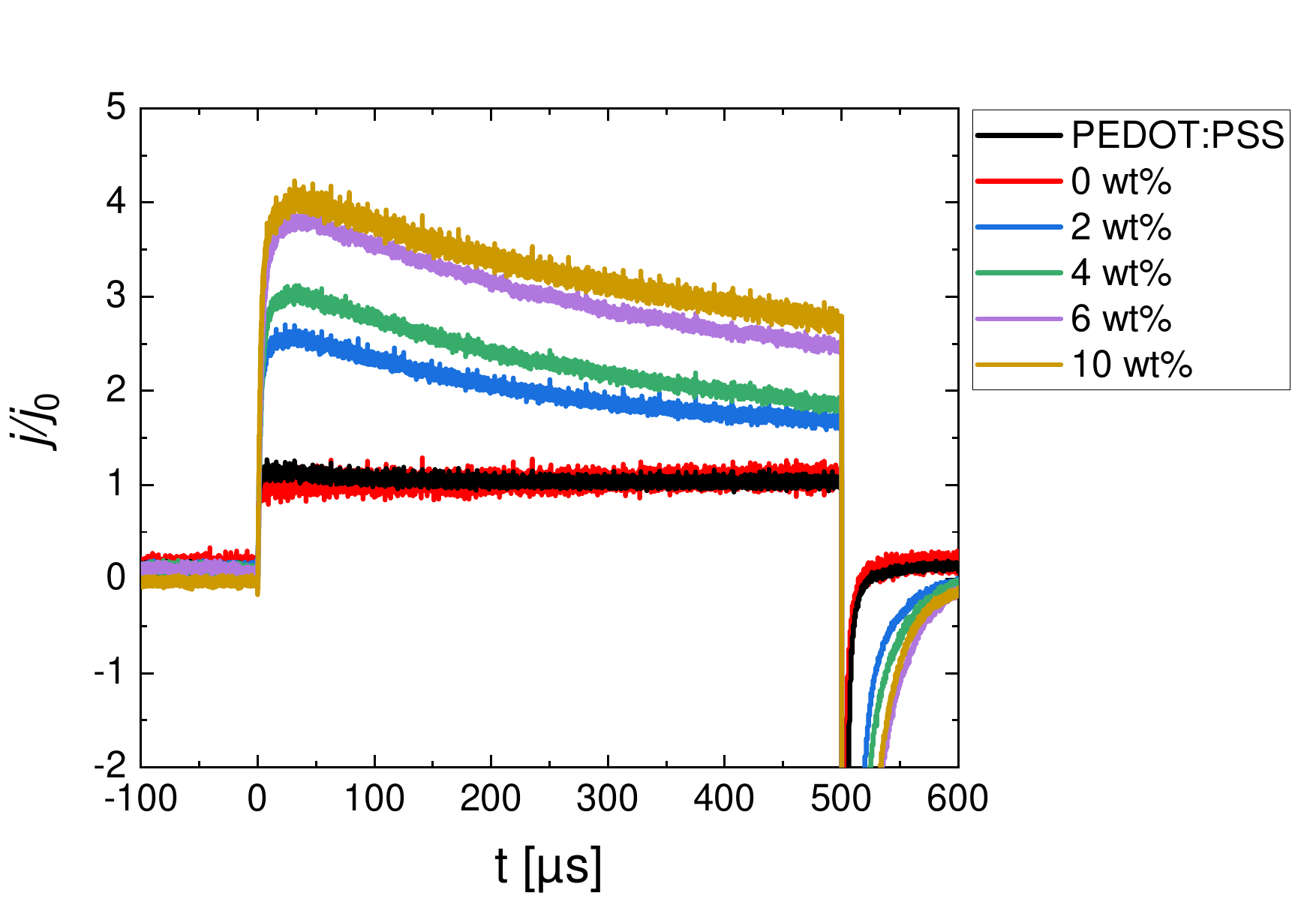}
\caption{CELIV transients for P3HT:PCBM solar cells with different HTLs.}
\label{fig:figS8}
\end{figure}

Exemplary CELIV~transients for solar cells with PEDOT:PSS or cross-linked P3HT at various dopant concentrations as HTL are shown in Figure~\ref{fig:figS8}. All measurements were performed at an  voltage pulse amplitude of~$\unit[2]{V}$, an offset voltage of~$V_\text{off} = \unit[0]{V}$ and a pulse length of~$\unit[500]{\mu s}$. The measured current is normalized to~$j_0$ and plotted versus time~$t$. For the samples with PEDOT:PSS and undoped P3HT, the transient current~$j(t)$ equals the current~$j_0$ given by the geometric capacitance of the device. In other words, both samples are effectively undoped. For the samples with a doped P3HT interlayer, there is an additional time dependent extraction current~$\Delta j(t)$ caused by the extraction of doping-induced charge carriers from the active layer.

\pagebreak

\begin{figure}
\includegraphics[width=0.7\textwidth]{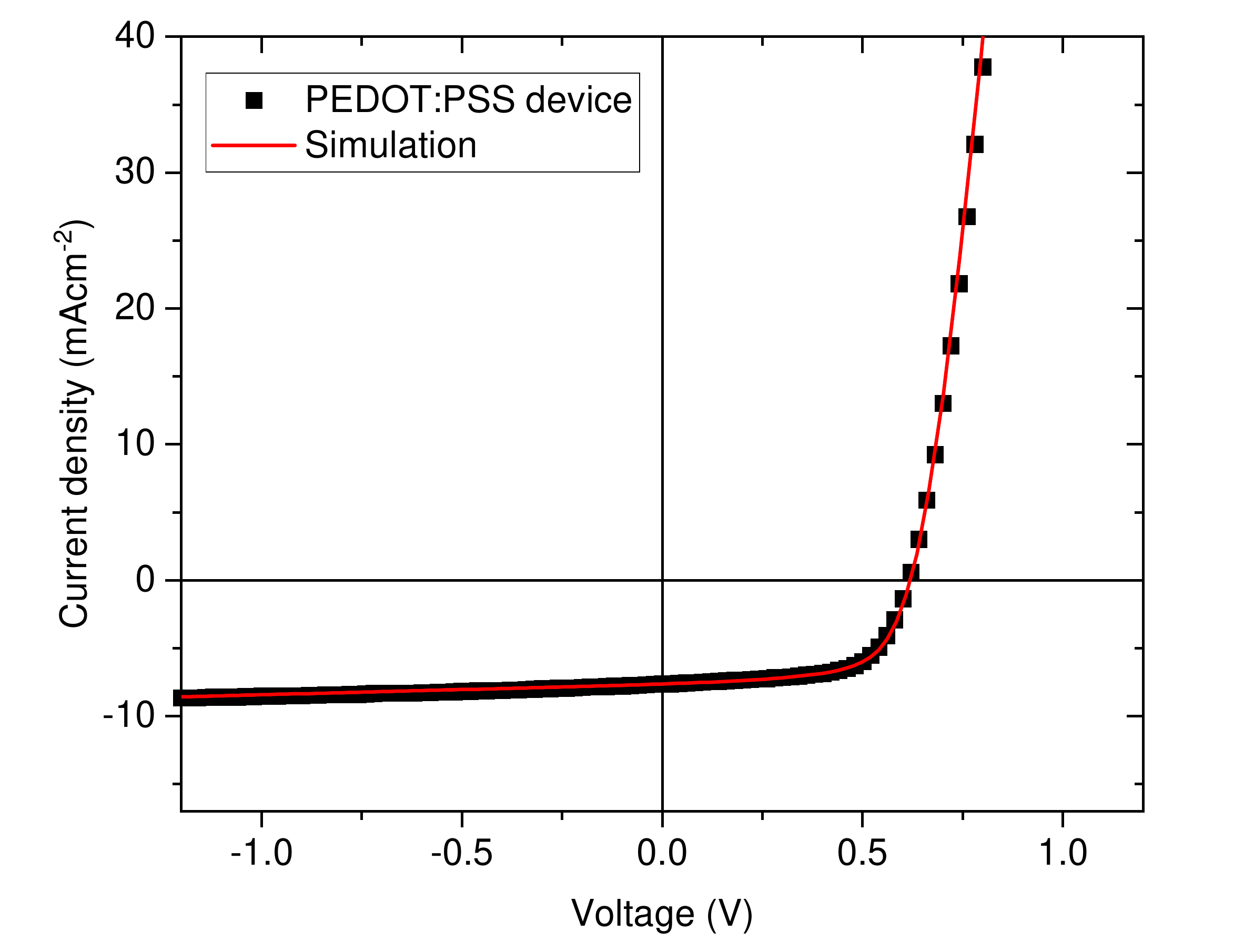}
\caption{Measured and modeled current--voltage curves for a P3HT:PCBM solar cell with a HTL of PEDOT:PSS. The parameters used in the numerical drift--diffusion model are given in Table~2 in the main text. The generation rate was taken from the transfer-matrix model, see Figure~\ref{fig:figS7}, and its spatial dependence was fully accounted for.}
\label{fig:figS9}
\end{figure}

\pagebreak

\begin{figure}
\includegraphics[width=0.7\textwidth]{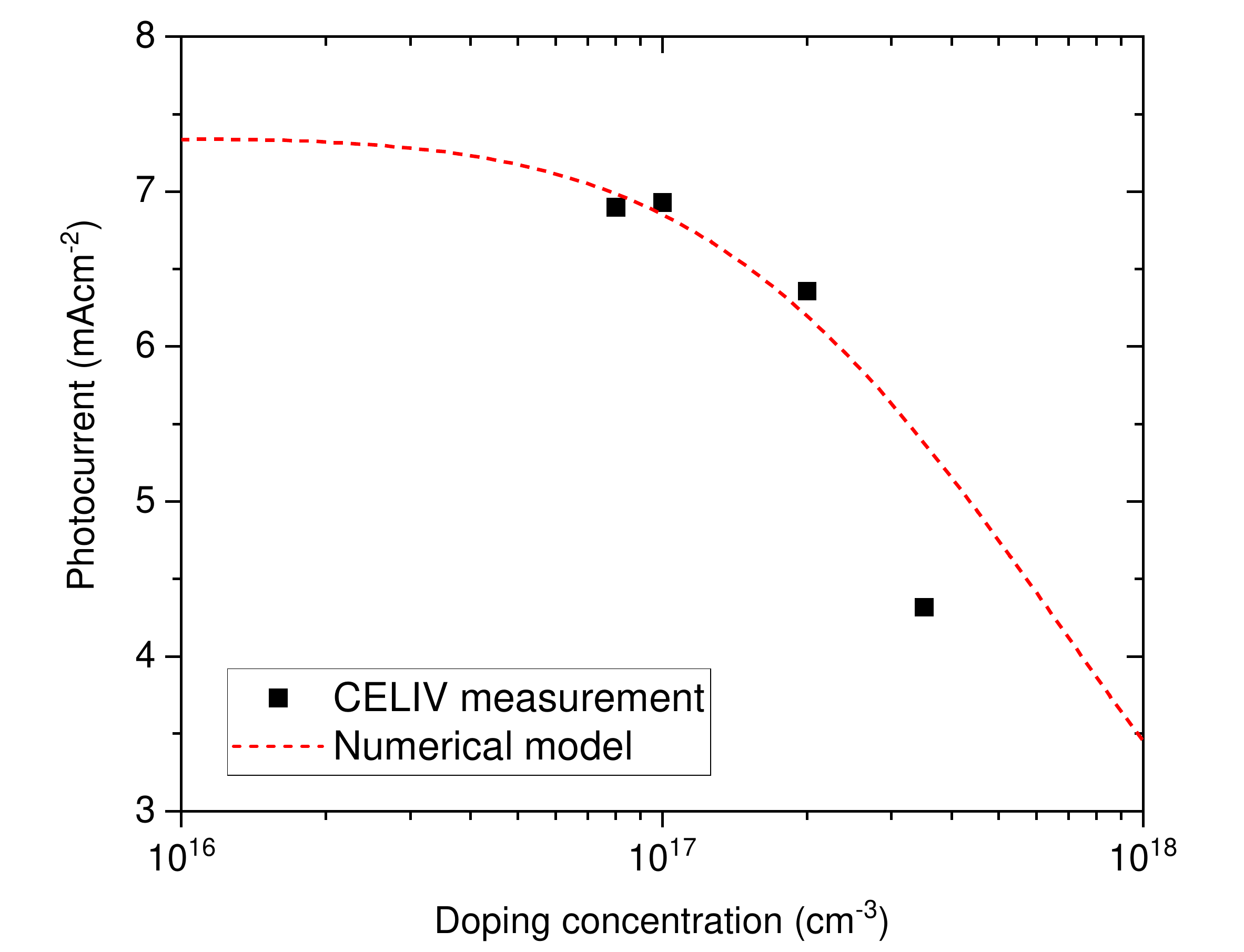}
\caption{Measured and modeled dependence of the photocurrent (shown is the short-circuit current~$\jsc$) on the doping concentration in the active layer. Experimental values were taken from the doping-CELIV measurements shown in the main text. The discrepancy between experiment and model at high doping concentrations suggest that the real doping profiles are not constant as assumed in the model, but show a more complex spatial dependence with an increased concentration towards the anode.}
\label{fig:figS10}
\end{figure}

\pagebreak
\bibliography{supplement}